 \documentclass[final,1p,times]{elsarticle}

\newcommand{\Msun}{\mbox{$M_{\odot}$}}
\newcommand{\Rsun}{\mbox{$R_{\odot}$}}

\newcommand{\kms}{\mbox{km s$^{-1}$}}
 \usepackage{graphics}
 \usepackage{graphicx}
 \usepackage{epsfig}

\usepackage{amssymb}
\usepackage{amsthm}

\usepackage{lineno}

\journal{New Astronomy}
\begin{document}
\begin{frontmatter}

  \title{Is NP\,Aqr a new near-contact binary?\tnoteref{t1}}
  \tnotetext[t1]{Based on observations collected at Catania Astrophysical Observatory (Italy)}
  \author[rvt]{C. \.{I}bano\v{g}lu}
  \ead{omur.cakirli@ege.edu.tr}
      \author[rvt]{\"{O}. ̃\c{C}ak{\i}rl{\i}\corref{cor1}}
      \author[rvt]{A. Dervi\c{s}o\v{g}lu}
          \cortext[cor1]{Corresponding author}
\address[rvt]{Ege University, Science Faculty, Department of Astronomy and Space Sciences, 35100 Bornova, \.{I}zmir, Turkey}
                                                                      
\begin{abstract}
We present radial velocities of the double-lined spectroscopic binary NP\,Aqr. The radial velocities
and the optical light curves obtained by Hipparcos and ASAS-3 are analyzed separately. The masses of the primary and
secondary components have been found to be 1.65$\pm$0.09 and 0.99$\pm$0.05 M$_{\odot}$, respectively. The cross-correlation 
functions indicate triple peaks which show presence of a tertiary star. The spectroscopic properties of this additional 
component resemble to that of the primary star. The analysis of the light curves yielded that the more massive primary star 
fills its corresponding Roche lobe. The secondary component is at or near Roche lobe indicating a new $\beta$ Lyrae-type 
near-contact binary. The orbital inclination is about 40$^{\circ}$ and, therefore, the observed light variations are produced 
only by the proximity effects. Due to the absence of eclipses, the effective temperature of the secondary star and the 
radii of the components could not be determined accurately. We conclude that NP\,Aqr is a non-eclipsing double-lined 
spectroscopic binary with a distance of about 134$\pm$7 pc. The absolute parameters of the components are also compared 
with the evolutionary models. While the location of the primary star seems to be suitable with respect to its mass in the
Hertzsprung-Russell diagram, the secondary component is located as if a star having a mass less than 0.6 M$_{\odot}$. This 
discrepancy is originated from very low effective temperature determined only from the light curve produced by proximity 
effects. The distance to the third star appears to be very close to that of the close binary which indicates that it may be 
dynamically bounded to the binary.
\end{abstract}
\end{frontmatter}

\linenumbers


\section{Introduction}
The light variability of NP\,Aquarii (=HD\,198528=HIP\,102935=NSV\,25363) was first suspected by Tobin, Viton \& Sivan (1994) 
during the Spacelab-1 Very Wide Field survey of UV-excess objects. This survey has revealed a variety of stellar objects with 
strong ultraviolet excess (see, for example, Viton et al. 1988). Perryman et al. (1997) confirmed its light variation 
using the H$_p$ magnitudes obtained by the Hipparcos mission. The type of the light variability could not be classified 
with the available data. Later on, Otero (2003) gave the epoch of minimum light, the orbital period and the variability 
class using the ASAS-3 (Pojmanski, 2002) and Hipparcos databases. He classified the star as a $\beta$ Lyrae-type eclipsing 
binary with an orbital period of 0.806982 day and with very shallow eclipses, namely 0.1 mag. Very recently, Kazarovets 
et al. (2006) designated the star HD\,198528 as NP\,Aqr using the criteria of {\it General Catalog of Variable Stars} in 
the 78$^{th}$ Name-list of Variable Stars. van Leeuwen (2007) re-examined the Hipparcos data and concluded that NP\,Aqr is 
an eclipsing binary with a spectral type of F0V at a distance of 186$\pm$27 pc. He also gives wide-band B-V and V-I color 
indices and interstellar reddening, E(B-V). Due to its very limited light variations and a spectral type of F0 
Handler (1999) included the star in 
the $\gamma$ Doradus candidates but with doubts that it may also be classified as a $\delta$ Scuti type variable. Handler 
and Shobbrook (2002) made Johnson  B, V and Cousins I$_c$  observations of the known and candidate $\gamma$ Dor stars to assess 
the relationship between the $\gamma$ Dor and $\delta$ Sct stars. They found slow variations with amplitudes of several 
hudredths of a magnitude. Taking into account a double-wave light curve they have excluded W UMa-type variability and 
suggested a possibility of an ellipsoidal variable with a period of about 0.807 day.

In this paper, we use the optical spectra of NP\,Aqr to reveal the nature of its light variability and physical properties 
combining with the photometric data obtained by Hipparcos and ASAS-3. The paper is organized as follows. In \S 2 the spectroscopic 
observations and data analysis are described. The spectroscopic and photometric results are combined and the absolute parameters 
of the stars are derived and the structure of NP\,Aqr is discussed in \S 3.  A brief conclusion is given in \S 4.  

\section{Spectroscopic observations}
Spectroscopic observations have been performed with the \'{e}chelle spectrograph (FRESCO) at the 91-cm
telescope of Catania Astrophysical Observatory. The spectrograph is fed by the telescope through an optical 
fibre ($UV$--$NIR$, 100 $\mu$m core diameter) and is located, in a stable position, in the room below the dome 
level. Spectra were recorded on a CCD camera equipped with a thinned back--illuminated SITe CCD of 
1k$\times$1k pixels (size 24$\times$24 $\mu$m). The cross-dispersed \'{e}chelle configuration yields a resolution 
of about 22\,000, as deduced from the full width at half maximum (FWHM) of the lines of the Th--Ar calibration lamp. The 
spectra cover the wavelength range from 4300 to 6700 {\AA}, split into 19 orders. In this spectral region, and 
in particular in the visual portion of the spectrum, there are several lines useful for the measurement of radial 
velocity, as well as for spectral classification of the stars.

The data reduction was performed by using the \'{e}chelle task of IRAF package following the standard steps: background 
subtraction, division by a flat field spectrum given by a halogen lamp, wavelength calibration using the emission lines of
a Th-Ar lamp, and normalization to the continuum through a polynomial fit. 

Sixteen spectra of NP\,Aqr were collected during the 20 observing nights between August 2$^{nd}$ and September 
23${^{th}}$, 2006. Typical exposure times for the NP\,Aqr spectroscopic observations were between 2600 and 
3000 s. The signal-to-noise ratios ($S/N$) achieved were between 70 and 115, depending on atmospheric condition. 

We also observed the radial and rotational velocity standards $\alpha$ Lyr (A0V), $\iota$ Psc (F7V), and 
50 Ser (F0V) with the same instrumentation. The average $S/N$ at continuum in the spectral region of interest 
was 150--200 for the standard stars. Cross-correlating the spectra of NP\,Aqr with standard stars usually yielded robust
correlation peaks with a FWHM of approximately 0.18 \AA. Fitting a Gaussian profile to these peaks allowed the center to 
be determined to an accuracy of about 0.02 \AA. This transforms to a radial velocity accuracy of $\sim$1 km s$^{-1}$. 

\begin{figure}
\includegraphics[width=13cm]{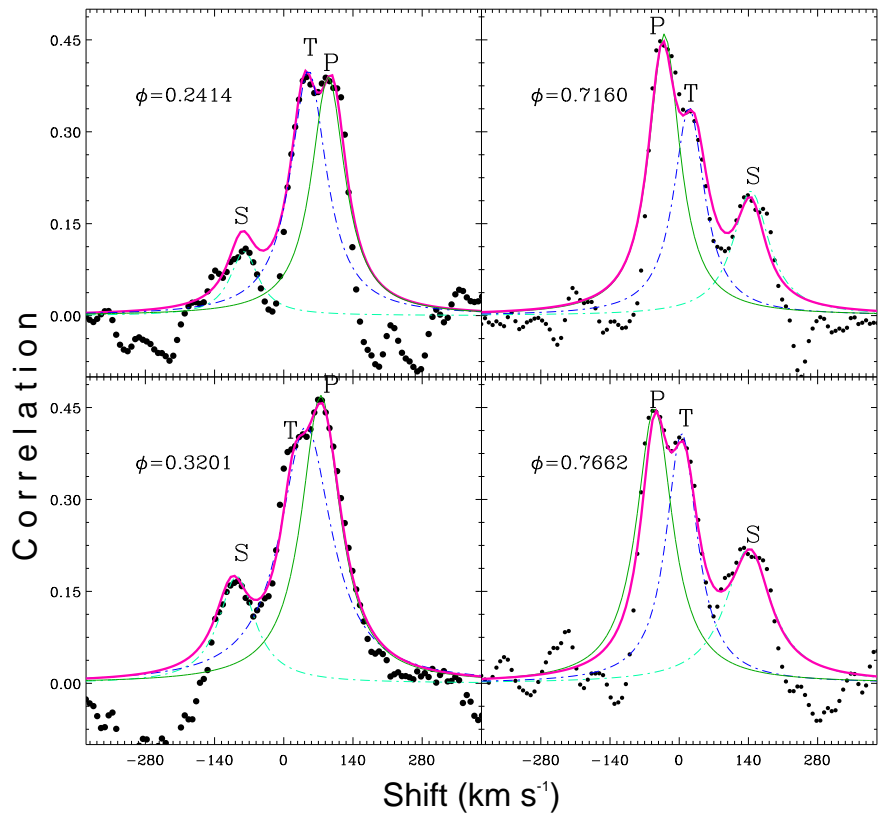}
\caption{Sample of Cross Correlation Functions between NP\,Aqr and the radial velocity template spectrum (50\,Ser) in
various orbital phases, especially at the maxima. The orbital phases given in each panel are computed with the 
ephemerides given by Otero (2003). The horizontal axis is relative radial velocities, and vertical axis is 
normalized cross-correlation amplitude. Note that 
splittings at stronger peaks. P, S and T refer to the primary, secondary and tertiary stars, respectively. }
\label{ccf;fig2}
\end{figure}

Double-lined spectroscopic binaries are characterized by the presence of two stellar spectra that appears in the 
cross-correlation function (CCF) as two peaks displacing back and forth according to the orbital motion of the 
system's components. The location of the two peaks allows the measurement of the radial velocity of each component 
at the phase of observation. The cross-correlation technique applied to digitized spectra is now one of the standard 
tools for the measurement of radial velocities in close binary systems. This helps to explore the binary mass-ratio 
distribution, especially in the low-mass regime. 

\begin{table}
\centering
\caption{Radial velocities of the components of NP\,Aqr. The columns give the heliocentric Julian date, the
orbital phase, the radial velocities of the two components and third star with the corresponding errors, and the average 
S/N of the spectrum. }
\begin{tabular}{@{}ccrcrcccccc@{}}
\hline
\textsf {HJD} & Phase& \multicolumn{2}{c}{Star 1 }& \multicolumn{2}{c}{Star 2 } & \multicolumn{2}{c}{Star 3 }& $<S/N>$ \\
  2\,453\,000+ &  & \textsf{{\bf V$_p$}} & $\sigma$ & \textsf{{\bf V$_s$}} & $\sigma$& \textsf{{\bf V$_t$}} & $\sigma$  \\
\hline
 53953.3919 &0.1410	&  80.0 &5.0  & -81.0  & 8.1   & 12.6    &  9.8  & 75     	  \\
 53953.5325 &0.3152	&  91.6 &3.0  & -94.4  & 8.4   & 26.4    &  8.8  & 88   	  \\
 53954.3766 &0.3612	&  79.0 &3.1  & -70.9  & 8.5   & 23.5    & 10.0  & 90	      \\
 53955.3771 &0.6011	& -27.9 &5.0  & 103.4  &14.5   & 35.1    & 11.0  & 87         \\
 53955.5104 &0.7662	& -54.0 &6.7  & 143.6  & 9.2   & 17.3    &  8.8  & 111$^a$	  \\
 53958.3998 &0.3467	&  84.0 &7.4  & -83.5  & 8.8   & 25.8    &  6.7  & 95$^a$     \\
 53970.4195 &0.2414	&  96.0 &5.2  &-106.4  & 6.8   & 27.6    &  7.1  & 106$^a$    \\
 53971.4214 &0.4830	&  27.4 &8.3  &  ---   & ---   &  ---    &  ---  & 80	      \\
 53973.3696 &0.8971	& -22.0 &8.0  &  99.2  &13.5   & 13.3    & 11.0  & 97         \\
 53975.3249 &0.3201	&  91.6 &2.4  & -90.0  & 7.8   & 38.4    &  7.9  & 115$^a$	  \\
 53980.3008 &0.4862	&  25.0 &7.1  &  ---   & ---   &  ---    &  ---  & 75         \\
 53981.2932 &0.7160	& -56.4 &5.6  & 148.1  & 7.7   & 13.9    &  8.0  & 106$^a$    \\
 53982.3099 &0.9759	&  10.0 &7.5  &  39.4  & 6.0   & ---     &  ---  & 60	      \\
 53983.2941 &0.1955	&  93.3 &5.9  & -99.5  & 9.1   & 26.4    &  9.9  & 81$^a$     \\
 53985.2766 &0.6522	& -42.2 &4.9  & 126.1  & 9.2   & 33.4    &  6.7  & 110$^a$    \\
\hline
\end{tabular}
\begin{list}{}{}
\item[$^a$]{\small Used also for rotational velocity ($v\sin i$) measurements.}
\end{list}
\end{table}

The radial velocity measurements of NP\,Aqr were obtained by cross--correlation of each \'{e}chelle order of NP\,Aqr spectra 
with the spectra of the bright radial velocity standard stars $\alpha$ Lyr (A0V), 59 Her (A3IV), $\iota$ Psc (F7V), and 
50\,Ser (F0V) whose radial velocities are $-13.5$ km\,s$^{-1}$, $-12.4$ km\,s$^{-1}$, +5.4 km\,s$^{-1}$, and -45.5 
km\,s$^{-1}$ respectively, determined by Nordstr\"om et al. (2004). For this purpose the IRAF task \textsf{fxcor} was used.

We applied the cross-correlation technique to several wavelength regions with well-defined absorption lines of the 
primary and secondary components. These regions include the following lines: Si\,{\sc iii} 4568 \AA, Mg\,{\sc ii} 
4481 \AA, He\,{\sc i} 5016 \AA, He\,{\sc i} 4917 \AA. The stronger CCF peak shows two peaks having nearly equal 
intensity and width, especially at both quadratures. One of these stronger CCF peaks should correspond to the more 
massive component that also has a larger contribution to the observed spectrum. To better evaluate the centroids of 
the peaks (i.e. the radial velocity difference between the target and the template), we adopted two separate Gaussian 
fits for the case of significant peak separation. However, we adopted three separate Gaussian fits for the case of small 
relative line shifts at near the conjunctions when spectral lines are particularly blended. In most cases, the Gaussian 
approximation gave reasonably good fits for the central parts of the CCFs.

Figure 1 shows examples of CCFs at various orbital phases. The two peaks, non-blended, correspond to each component 
of NP\,Aqr. The stronger peak in the CCFs corresponds to the more massive component, which is also the more luminous at 
the observational wavelengths. The CCFs shown in Fig. 1 reveal presence of at least three peaks in the spectra of 
NP\,Aqr, corresponding to phases near the quadratures. However, the spectral lines of the tertiary component appear to 
be blended with those of the primary star. The peak between the peaks of primary and secondary component could be 
related to a possible tertiary component. It should be noted that this peak can be resolved for very limited 
spectra taken at both quadratures. The measured radial velocities for the components of close binary and for 
the third star are given in Table 1. However, given the noise in the CCFs, we cannot exclude the possibility that 
this feature is a spurious peak. We estimate roughly the light contribution of the primary, secondary and tertiary 
components in the region covered by the V bandpass as 0.44$\pm$0.11, 0.16$\pm$0.04 and 0.40$\pm$0.07, respectively, 
using the FWHM of the CCFs. 
   
\subsection{Analysis of the radial velocities}
The radial velocity measurements, listed in Table 1 together with their standard errors, are weighted means of the 
individual values deduced from each order (see, e.g., Frasca et al. 2006). The observational points and their error 
bars are displayed in Figure 2 as a function of orbital phase, computed using the ephemeris given by Otero (2003), 

Min\,I\,=\,JDH\,24\,47985.661+0.806982\,E.

We also plot the radial velocities of the third star in Figure 2 (asterisks). Since the spectral lines of the 
primary star and the tertiary component are blended we could measure the radial velocities of the third star 
using the peaks of CCFs. For this reason the RVs of the third star appear to vary with respect to the orbital 
phases of the binary. The solution of RVs of NP\,Aqr yields the semi-amplitudes of the more massive, 
and less massive components as K$_1$=76$\pm$2 km s$^{-1}$ and K$_2$=127$\pm$2 km s$^{-1}$, respectively; 
systemic velocity of V$_{\gamma}$=21$\pm$1 km s$^{-1}$ and an eccentricity of nearly zero, indicating a circular orbit.

\subsection{Rotational velocities}
The width of the cross-correlation profile is a good tool for the measurement of $v \sin i$ (see, e.g., 
Queloz et al. 1998). The projected rotational velocities ($v \sin i$) of the two components were obtained by 
measuring the FWHM of the CCF peaks in nine high-S/N spectra of NP\,Aqr acquired at phases, where the 
spectral lines have the largest Doppler-shifts. In order to construct a calibration curve FWHM--$v \sin i$, we 
have used an average spectrum of $\iota$ Psc, acquired with the same instrumentation. Since the rotational 
velocity of $\iota$ Psc is very low but not zero ($v \sin i$ $\simeq$11 km s$^{-1}$, e.g., Royer, Zorec \& Fremat 
2004 and references therein), it could be considered as a useful template for A-type stars rotating faster 
than $v \sin i$ $\simeq$ 10 km s$^{-1}$. The spectrum of $\iota$ Psc was synthetically broadened by convolution 
with rotational profiles of increasing $v \sin i$ in steps of 5 km s$^{-1}$ and the cross-correlation with 
the original one was performed at each step. The FWHM of the CCF peak was measured and the FWHM-$v \sin i$ 
calibration was established. The $v \sin i$ values of the two components of NP\,Aqr were derived from the FWHM of 
their CCF peak and the aforementioned calibration relations. The FWHM values were derived for a few wavelength 
regions and for the best spectra obtained near the quadratures. The mean projected rotational velocities of the primary, 
secondary and tertiary components are found to be 72$\pm$7 km s$^{-1}$, 63$\pm$10 km s$^{-1}$ and 74$\pm$11 km s$^{-1}$, 
respectively.

\subsection{Spectral classification}
We have used our spectra to classify the primary component of NP\,Aqr. For this purpose we have degraded the spectral
resolution from 20\,000 to 3\,000, by convolving them with a Gaussian kernel of the appropriate width, and we have 
measured the equivalent width ($EW$) of photospheric absorption lines useful for the spectral classification. We 
have followed the procedures of Hern\'andez et al. (2004), choosing hydrogen and helium lines in the blue-wavelength 
region, where the contribution of the secondary component to the observed spectrum is negligible. From several spectra 
we measured  $EW_{\rm H\gamma}=11.3\pm 0.9$\,\AA, $EW_{\rm H\alpha}=9.79\pm 0.11$\,\AA, $EW_{\rm H\beta}=10.79\pm 0.13$\,\AA, 
and $EW_{\rm MgII\lambda 4481}=0.35\pm 0.04$\,\AA.

From the calibration relations $EW$--Spectral-type of Hern\'andez et al. (2004), we have derived a spectral type F1V 
with an uncertainty of about 1 spectral subclass. The effective temperature deduced from the calibrations of Drilling 
\& Landolt (2000) or de Jager \& Nieuwenhuijzen (1987) is about 6\,900\,K (F2V). The spectral-type uncertainty leads 
to a temperature error of $\Delta T_{\rm eff} \approx 400$\,K. 

The B-V and V-I colors and the reddening in B-V are given by van Leeuwen (2007) as 0.348$\pm$0.015, and 0.410$\pm$0.020 and 0.015 
mag, respectively. Using the color-temperature relation given by Drilling \& Landolt (2000) we estimate an effective temperature 
of 7\,100$\pm$90\,K  for the primary component of NP\,Aqr. However, we find 6\,850$\pm$80\,K, 7\,000$\pm$100\,K, 7\,060$\pm$100\,K 
for the same colors using the calibrations given by Popper (1980), Alonso, Arribas \& Martinez-Roger (1996) and Flower (1996), respectively. On 
the other hand, the $y/V$, $b-y$, $m_1$, $c_1$, and $\beta$ values are given by Tobin, Viton \& Sivan (1994) 
and Hauck \& Mermillod (1998) as 7.640$\pm$0.034,  0.201$\pm$0.006, 0.166$\pm$0.006, 0.800$\pm$0.003 and 2.775, respectively. 
The calibration of Drilling and Landolt (2000) yields an effective temperature of 7\,170$\pm$55\,K. However, the infrared colors 
given by Cutri et al. (2003) are J-H=0.130$\pm$0.040 and H-K=0.088$\pm$0.039 mag which correspond to a temperature of 7\,100$\pm$100\,K according to the calibrations of Tokunaga (2000). The temperature uncertainty of the primary component 
results from considerations of spectral type uncertainties and temperature calibration differences. The weighted mean of 
the effective temperature of the primary star is 7\,050$\pm$95 K.

\begin{figure*}
\includegraphics[width=14cm]{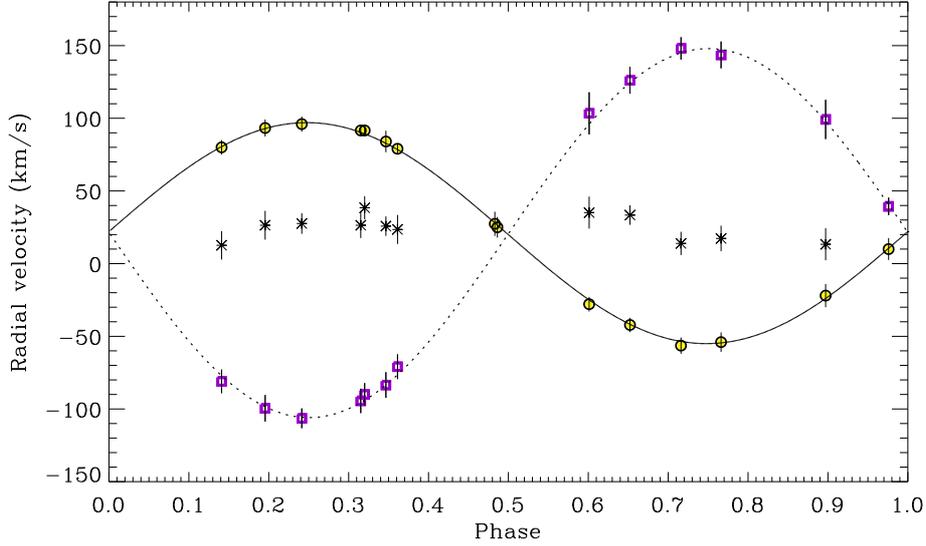}
\caption{The observed radial velocities of NP Aqr as a function of orbital phase, with the velocity curves corresponding 
to the adopted orbital elements drawn through them. Open circles and squares represent the velocities of the primary and 
secondary components, respectively, and the asterisks represent the velocities of the tertiary component. The vertical 
lines on the velocities indicate their error bars. }
\label{RV;fig4}
\end{figure*}

\subsection{Orbital solution}
The first photometric observations of NP\,Aqr were made by the Hipparcos mission and 49 H$_p$ magnitudes were listed by 
van Leeuwen (2007). These magnitudes were obtained in a time interval of about three years. The accuracy of the Hipparcos 
data is about $\sigma_{H_p}$ $\sim$ 0.01. The peak-to-peak amplitude 
of the light variation is about 0.1 mag. These measurements are plotted against the orbital phase in the bottom panel of 
Fig. 3.

NP\,Aqr was identified in the All-Sky Automated Survey (ASAS, Pojmanski 2004) as a detached eclipsing binary system with 
a maximum, out-of-eclipse $V$-bandpass magnitude of 7.59$^{\rm m}$ and a period of 0.8070 day. The light curve seems 
to exhibit periodic eclipses with depths of primary and secondary eclipse of $\sim 0^{\rm m}.1$ and $\sim 0^{\rm m}.5$, 
respectively. The scatter of the data in the out-of-eclipse phases was about $0^{\rm m}.03$. We plot the ASAS data in 
the top panel of Fig. 3. It is clear that both light curves are very similar but the Hipparcos magnitudes are 
smaller about 0.2 mag than those of the ASAS. Since the number of the individual observations obtained by Hipparcos mission and 
by ASAS are not too much, we use all of them for the analysis of the light curves, separately. We used the last 
version of the Wilson-Devinney code (Wilson and Devinney 1971) in order to obtain the 
inclination of the orbit, the fractional radii and luminosities of the components as well as the effective 
temperature of the cooler component. In the light curve solution we keep some parameters fixed whose values can 
be estimated from radial velocities and global stellar properties. The limb-darkening coefficients were estimated 
from van Hamme (1993), the gravity-darkening exponents were assumed to be $g_1$=1.0 and $g_2$=0.32, according to 
the von Zeipel's law, because the secondary component seems to have convective envelope. The bolometric albedos 
were set to $A_1$=1.0 and $A_2$=0.5 (Lucy 1967), suitable for the radiative and convective envelopes. The mass 
ratio, the key parameter in the solution, is taken to be 0.6 adopted from the radial velocity analysis. 

We started the light curve analysis with effective temperature of 7\,050\,K for the primary component. Initially, 
the Mode-2 of the Wilson-Devinney code, referring to the detached Algols, was adopted for the analysis of the 
light curves. The adjustable parameters in the analysis were the inclination of the orbit ($i$), the surface 
potentials($\Omega_1$, $\Omega_2$), and the effective temperature of the secondary (T$_2$), the luminosity of 
the primary star (L$_1$) and light contribution of the tertiary component ($l_3$). Using trial-and-error method 
we could not obtain a set of parameters, which marginally represented the observed light curves. So, we kept the 
orbital inclination as fixed, starting from 38 to 45 degrees. The iterations were carried out automatically until 
convergence, and a solution was defined as the set of parameters for which the differential corrections were smaller 
than the probable errors. However, after a few runs we found that the primary star's volume exceeded its inner Roche 
lobe whereas the secondary star was inside its lobe. So, we arrived at a decision that NP\,Aqr is not a detached 
system. Next, we tried $Mode-3$ (for contact systems), $Mode-4$ (primary star fills its lobe) and $Mode-5$ (secondary 
star fills its lobe). The fits of the computed light curves obtained with Mode-4 to the observations at an inclination 
of 40$^{\circ}$ are satisfactory. The smallest sum of residuals squared was also arrived at an inclination of 40$^{\circ}$. Since 
spectroscopic observations revealed signs of a tertiary component which contributes an amount of light as large as the 
primary component we also took the value of ($l_3$) as an adjustable parameter. The value of $l_3$ in units of total 
triple star system light was estimated to be 0.08, too small compared with the estimate from spectra. If we increase 
the inclination of the orbit to 41$^{\circ}$ the eclipses set in and the light contribution of the third star can be better 
determined. When the orbital inclination of the binary is taken as 44$^{\circ}$ the light contribution of the third star 
reaches to 0.20, almost half of the value obtained from the spectra. However, photometric light curves obtained by 
Hipparcos and ASAS do not clearly show evidence of the eclipses. So, we conclude that the observed low-amplitude 
light variations are produced by the proximity effects rather than eclipses. Such a low-amplitude light variation 
may give unreliable solutions. Therefore, the rough values of the orbital parameters were estimated in this study, 
whereas determination of the $l_3$ from the available light curves appears to almost impossible.             

The final orbital and stellar parameters from the simultaneous light curve analysis are listed in Table 3. The 
parameters that were adjusted have standard errors as computed by the program. The computed light curve is 
compared with the observations in Fig. 3. Since only the light variations originated from the proximity effects 
could be observed, the real errors of the parameters should be larger than those computed by the WD code. 
Furthermore, it should also be noted that the presence of a tertiary component slightly affects the fractional 
luminosities of the components. 
  
The results of the light curve analysis indicate that NP\,Aqr is very similar to that so-called inverse Algols. While 
the primary star appears to filling its Roche lobe the secondary star is very close its lobe, indicating a near 
contact binary (NCB). This is the case for most NCBs whose light curves are of the $\beta$-Lyrae type. NCBs are 
distinct from the contact binaries with no large-scale energy transport from one component to the other. Shaw 
(1994) divided the NCBs into two sub-groups designated by the so-called prototypes FO\,Vir and V1010\,Oph. The 
stars belong to the latter subclass, i.e. V1010\,Oph-type NCBs, are defined as those with a normal primary at 
or near the Roche lobe and a secondary that is up to 1.5 times oversized, but well inside the Roche lobe. In 
Figure 4 we show the volumes of the components and their corresponding Roche lobes. While the more massive 
primary star is filling its Roche lobe the secondary is close to its lobe. Recently, Oh (2005) collected physical 
parameters of all the NCBs. He proposed a relationship between the mass ratio and luminosity ratio of the NCBs as 
(L$_2$/L$_1$)=(M$_2$/M$_1$)$^{1.45}$. We estimate that the secondary star has a radius of 1.67 times larger with 
respect to its mass. The light ratio determined from the spectra is about 0.398 which is smaller only 20 per cent 
than that computed from the mass-to-luminosity ratios relationship. These results lead to the conclusion that 
NP\,Aqr fulfills almost all properties of the V1010\,Oph-type NCBs and is very similar to the NCB EU\,Hya.

\begin{figure*}
\includegraphics[width=10cm]{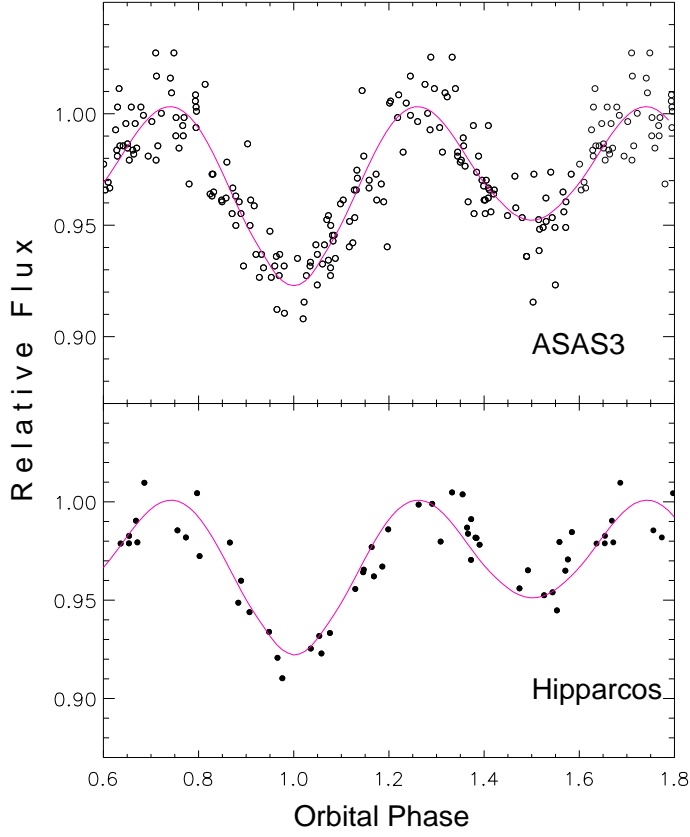}
\caption{The light curves of NP Aqr obtained by Hipparcos (bottom panel) and ASAS-3 (top panel). The light curve corresponding to 
the adopted orbital elements is shown by continuous line.}
\label{RV;fig4}
\end{figure*}

\begin{figure*}
\includegraphics[width=10cm]{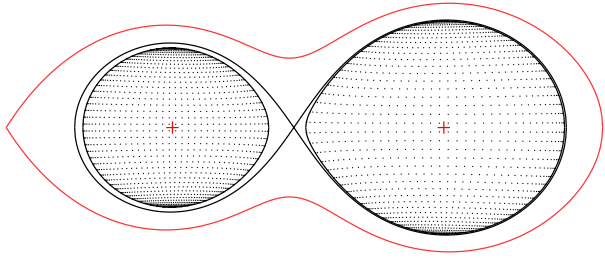}
\caption{The three-dimensional configuration of the components of the system in the orbital plane according to our solution.}
\label{RV;fig4}
\end{figure*}

\begin{table}
\caption{Orbital parameters for NP Aqr, obtained from the RVs analysis.}
\label{tab: orbit}
\begin{tabular}{lc}
\hline
$P_{orb}$ (days)		&0.806982 			\\
V$_{\gamma}$ (\kms) 	&21$\pm$1           \\
$K_1$ (\kms)			&76 $\pm$2          \\
$K_2$ (\kms)			&127 $\pm$2         \\
$a_1\sin i$ (\Rsun)		&1.212 $\pm$0.032	\\ 
$a_2\sin i$ (\Rsun)		& 2.025 $\pm$0.037	\\
$q\ (=M_2/M_1)$		 	&0.598$\pm$0.018	\\
$M_1\sin^3 i$ (\Msun)	&0.438$\pm$0.018	\\
$M_2\sin^3 i$ (\Msun)	&0.262$\pm$0.013	\\		
\hline 
\end{tabular}
\end{table}

\begin{table*}
\caption{Results of the ASAS-3 and Hipparcos light curves analysis for NP\,Aqr. The adopted 
values are the weighted means of the values determined from the individual light curves.}
\begin{tabular}{lc}
\hline
Parameters & ASAS3   \\
\hline
$i^{o}$			         &40\,[Fix]	 	 	\\
T$_{eff_1}$ (K)		     &7\,050[Fix]	 	\\
T$_{eff_2}$ (K)			 &4250$\pm$385	 	\\
%$\Omega_1$				 &3.0634$\pm$		\\
$\Omega_2$				 &0.306$\pm$0.012	\\ 			    
r$_1$					 &0.423$\pm$0.013   \\
r$_2$					 &0.334$\pm$0.020 	\\
${L_{1}}/{(L_{1}+L_{2})}$&0.95$\pm$0.04 	\\
${L_{2}}/{(L_{1}+L_{2})}$&0.05$\pm$0.03 	\\
$\chi^2$				 &0.049				\\  			    
\hline
\end{tabular}
\end{table*}

\begin{table}
 \setlength{\tabcolsep}{2.5pt} 
  \caption{Fundamental parameters of the system.}
  \label{parameters}
  \begin{tabular}{lcc}
  \hline
%  & \multicolumn{2}{c}{NSVS06507557} 		\\
   Parameter 						& Primary		&		Secondary	\\
   \hline
   Mass (M$_{\odot}$) 				&1.65 $\pm$0.09 &0.99 $\pm$0.05	\\
   Radius (R$_{\odot}$) 			&2.13 $\pm$0.09 &1.67 $\pm$0.07	\\
   $\log~g$ ($cgs$) 				&4.00 $\pm$0.02 &3.99 $\pm$0.02	\\
   $T_{eff}$ (K)					&7050 $\pm$95	&4250 $\pm$385 	\\
   $\log~(L/L_{\odot})$				&1.00 $\pm$0.04      &-0.08 $\pm$0.16     	\\
   $(vsin~i)_{calc.}$ (km s$^{-1}$)	&86   $\pm$1  	 &68 $\pm$2	 	\\
   $(vsin~i)_{obs.}$ (km s$^{-1}$)	&72   $\pm$7  	 &63 $\pm$10	 	\\
   $d$ (pc)							& \multicolumn{2}{c}{134$\pm$7}	\\
\hline  
  \end{tabular}
\end{table}

\section{Discussion }
Combining the parameters in Tables 2 and 3, we derived the astrophysical parameters of the components and other 
properties in Table 4. Luminosities are computed directly from the radii and effective temperatures of the 
components. While the light contribution of the secondary component to the total light of the binary is estimated 
from the spectra to about 0.27, the contribution is derived from the light curve analysis as low as 0.05. As 
we mentioned in the previous section neither the temperature nor the light contribution of the secondary component 
are estimated accurately because of the lower orbital inclination, e.g. absence of eclipses. We compared the 
positions of the components of NP\,Aqr in the $log~T_{eff}$-$log~L/L_{\odot}$ diagram with the evolutionary 
tracks of Girardi et al. (2002), as shown in Fig. 5. The absolute parameters of the primary star place it very 
close to the evolutionary track of a 1.65 M$_{\odot}$ single star with solar abundance. However, the secondary 
component appears to have lower luminosity and lower effective temperature with respect to its mass. As we noted 
in the section dealing with the light curve analysis the observed light variation of NP\,Aqr is produced only 
from the proximity effects. Due to the lower orbital inclination eclipse phenomena are not produced. In addition, 
presence of a third component also lightens the effect originated from the proximity of close binary. Therefore, 
the analysis of the observed light curves can not yield the accurate parameters of the binary. We do not want 
to make further speculation about the location of the less massive component in the HR diagram and its evolutionary 
status, due to the very poor estimation of its parameters.  

\begin{figure*}
\includegraphics[width=10cm]{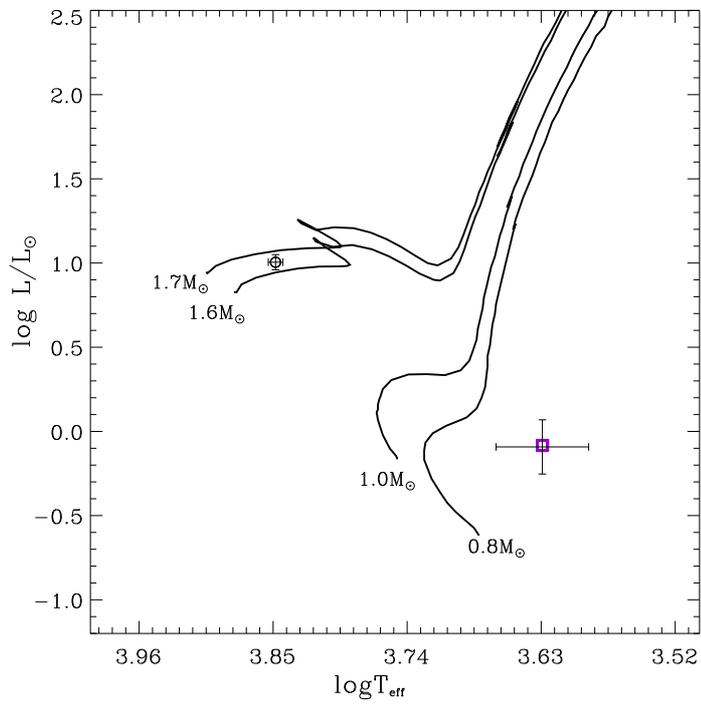}
\caption{Comparison between evolutionary models and the physical parameters of NP\,Aqr in the $log~T_{eff}$ - $log~L/L_{\odot}$ 
diagram. Theoretical evolutionary tracks with $Z=Z_{\odot}$ and masses of 0.8, 1.0, 1.6 and 1.7 \Msun~ are adopted from 
Girardi et al. (2000). The circle and square denote the primary and secondary stars, respectively.}
\label{RV;fig4}
\end{figure*}

NP\,Aqr was measured by the Hipparcos mission, but the published parallax is rather uncertain. In the first (Perryman et 
al.1997) and the last (van Leeuwen 2007) versions of the Hipparcos catalogs the parallaxes are given as 
$\pi_{Hip}$=5.50$\pm$1.31 and 5.39$\pm$0.76 mas, respectively. The latter corresponds to a formal distance of 186$\pm$27 
pc. Based on the BVIJHK magnitudes of the binary, cleaned from the contribution of third body, interstellar extinction 
given by Tobin, Viton \& Sivan (1994) and the bolometric corrections given by Girardi et al. (2002) we estimate a weighted 
distance to the spectroscopic binary NP\,Aqr to be 134$\pm$7 pc. The distance to the system determined in this 
study has very low uncertainty when compared with that measured by Hipparcos mission. The third star has 
characteristics of an F1-2 main-sequence star, similar to the primary component of the spectroscopic close 
binary, which would give it an absolute magnitude of about 3.2$\pm$0.5 mag and a distance of about 122$\pm$28 pc, 
if it is a single star. This result indicates that the third star would plausibly be dynamically bounded to the 
close binary. New spectra of the NP\,Aqr with the higher S/N ratio and resolution are needed to clarify the nature 
of the third star.

\section{Conclusion }
The first spectoscopic observations of the relatively short-period close binary NP\,Aqr are presented. CCFs 
indicate that NP\,Aqr is a triple system. The radial velocities of both components are obtained and analyzed for the 
spectroscopic elements. NP\,Aqr has been observed also by the Hipparcos mission and ASAS-3 which yielded almost complete 
light curves of the system. Both light curves are analyzed separately and the resultant orbital elements are determined. 
The analysis of the light curves revealed that the light variation with an amount of about 0.1 mag is produced from the  
proximity effects. In other words, NP\,Aqr is a double-lined, non-eclipsing spectroscopic binary. The radii of the 
components and effective temperature of the secondary star could be determined with great uncertainty due to the 
absence of the eclipses and also presence of a third star. Comparing stellar evolutionary models for the mass of 
the components we find that location of the primary star in the HR diagram is suitable with its mass. However, 
the secondary star looks like a low-mass star. While the radius of this star is as large as 1.67 times with respect 
to its mass, the effective temperature is unexpectedly low. The effective temperatures of the binary star's components 
can be determined by comparing the depths of the eclipses. Since the stars of NP\,Aqr do not show eclipses due to the small 
orbital inclination, the eclipses do not set in. Our analysis indicates that the star NP\,Aqr shows most of the 
characteristics of the so-called near-contact binaries. Moreover, we detected a third component in the system NP\,Aqr. 
The third star looks like the primary component in the spectra. Using this property and assuming that it is a 
main-sequence star we determined its distance to be 122 pc, at nearly the same distance with the spectroscopic binary.

\section*{Acknowledgments}
We thank Prof.\ G.\ Strazzulla, director of the Catania Astrophysical Observatory, and Dr. G.\ Leto, responsible 
for the M. G. Fracastoro observing station for their warm hospitality and allowance of telescope time for the 
observations. In addition, \"{O}\c{C} is grateful to all the people working at the Catania Astrophysical Observatory 
for creating a stimulating and enjoyable atmosphere and, in particular, to the technical staff of the OAC, namely 
P.\ Bruno, G.\ Carbonaro, A.\ Distefano, M.\ Miraglia, A.\ Miccich\`e, and G.\ Occhipinti, for the valuable support 
in carrying out the observations. EB{\.I}LTEM Ege University Science Foundation Project No:08/B\.{I}L/0.27 and Turkish 
Scientific and Technical Research Council for supporting this work through grant Nr. 108T210. This research has been also 
partially supported by INAF and Italian MIUR. This research has been made use of the ADS and CDS databases, operated at 
the CDS, Strasbourg, France.

\end{document}